# Parametric dependence of hot electron relaxation timescales on electron-electron and electron-phonon interaction strengths

*Richard B. Wilson*[1,2,*], *Sinisa Coh*[1,2,*]

1) Materials Science and Engineering, University of California - Riverside, CA 92521, USA

2) Mechanical Engineering, University of California - Riverside, CA 92521, USA

* rwilson@ucr.edu and sinisacoh@gmail.com.

Understanding how photoexcited electron dynamics depend on electron-electron (e-e) and electron-phonon (e-p) interaction strengths is important for many fields, e.g. ultrafast magnetism, photocatalysis, plasmonics, and others. Here, we report simple expressions that capture the interplay of e-e and e-p interactions on electron distribution relaxation times. We observe a dependence of the dynamics on e-e and e-p interaction strengths that is universal to most metals and is also counterintuitive. While only e-p interactions reduce the total energy stored by excited electrons, the time for energy to leave the electronic subsystem also depends on e-e interaction strengths because e-e interactions increase the number of electrons emitting phonons. The effect of e-e interactions on energy-relaxation is largest in metals with strong e-p interactions. Finally, the time high energy electron states remain occupied depends only on the strength of e-e interactions, even if e-p scattering rates are much greater than e-e scattering rates.

**Introduction**

Absorption of light by a metal generates a nonthermal distribution of electrons and holes [1-3]. In the femtoseconds to picoseconds following absorption, a complex cascade process emerges from individual electron-electron (e-e) and electron-phonon (e-p) scattering events [4-6]. This cascade process drives the system into a new equilibrium state.

We characterize the emergent nonequilibrium electron cascade process with two time-scales, $\tau_H$ and $\tau_E$. Time $\tau_H$ measures how long the metal contains highly excited electrons with energy comparable to that of the incoming photons, $hv$. Somewhat arbitrarily, we define $\tau_H$ as the time for the number of highly excited electrons with energy greater than or equal to $hv/2$ to drop by a factor of $1/e$, see Figure 1. Another emergent time scale shown in Figure 1 is $\tau_E$. Time $\tau_E$ is the time required for the total energy stored by all nonequilibrium electrons to drop by a factor of $1/e$.

Time-scales $\tau_E$ and $\tau_H$ are critical, and distinct, figures of merit for a variety of scientific and engineering endeavors, such as photocatalysis, ultrafast magnetism, and others. Ultrafast magnetic phenomena are commonly driven by $\tau_E$ because they depend on how quickly spatial gradients in internal energy are relaxed [7-13]. On time-scales shorter than $\tau_E$, nonequilibrium electrons transport energy at rates that are 1-2 orders of magnitude faster than is possible after electrons and phonons thermalize [11,14-17]. On the other hand, several recent studies suggest photocatalytic performance of plasmonic metal nanoparticles is governed by $\tau_H$ [18-21]. High energy electrons are hypothesized to drive chemical reactions [18-21]. However, this hypothesis remains controversial because it is difficult to differentiate the effect of temperature rises from the effect of high energy nonequilibrium electrons [22,23].

The fundamental importance of electron dynamics has motivated extensive theoretical [4,9,16,18,19,24-29] and experimental [2,5,6,30-39] study of relaxation times like $\tau_E$ and $\tau_H$. These prior studies provide descriptions of how nonequilibrium electron distributions in specific materials like Au, Al, and Cu evolve as a function of time [5,24,25,30-32,40]. Early work by Tas and Maris [5] and Groeneveld *et al.* [30] found the e-e scattering increases the rate of energy transfer to the lattice by increasing the number of excitations. Time-resolved two-photon photoemission studies have found e-e interactions cause high energy electrons to decay on time-scales of tens of femtoseconds after photoexcitation [33]. Mueller and Rethfeld provided a detailed analysis of how various aspects of the collision integrals and rate-equations effect nonequilibrium electron dynamics in Au, Al, and Ni [25]. Other studies have integrated first-principles calculations of band-structure [41], photon absorption [18,19,42], and e-p interactions [18,19] into models for nonequilibrium electron dynamics to improve agreement with experiment. Recent work by the plasmonics community has focused on understanding how hot electrons effect photocatalytic efficiencies in plasmonic systems [3,22,23,43-47].

Surprisingly, no systematic study of how $\tau_E$ and $\tau_H$ depends on e-e vs. e-p interactions exists. As a result, significant confusion persists regarding the best method for estimating $\tau_E$ and $\tau_H$ from material properties such as quasi-particle lifetimes. Estimates in the literature for the energy relaxation time $\tau_E$ of various metals[27] almost always underestimate the importance of e-e interactions [5,24,25,30,32,40]. By far the most common method for estimating $\tau_E$ of a metal is the two-temperature model [27,29,48]. The two-temperature model neglects nonthermal effects, and therefore neglects the important role of e-e interactions. Alternatively, the relaxation time of high energy electrons $\tau_H$ is often incorrectly estimated from a simplified Boltzmann rate equation with a Matthiessen's-like rule [45,49-51], resulting in $\tau_H^{-1} \approx \tau_{ee}^{-1} + \tau_{ep}^{-1}$. Here $\tau_{ep}$ is the electron-phonon quasi-particle scattering time. This treatment leads to the incorrect conclusion that, since e-p scattering rates are stronger than e-e scattering rates, $\tau_H$ depends on the strength of e-p interactions. In other-words, the Matthiessen's rule estimate for the relaxation times of a nonequilibrium electron distribution will dramatically overestimate the importance of e-p interactions.

Here, we present calculations of the dynamics of photoexcited electrons to quantify how $\tau_E$ and $\tau_H$ depend on electron-electron (e-e) and electron-phonon (e-p) interaction strengths. In contrast to the two-temperature model prediction of $\tau_E = \gamma_{ep}^{-1}$, we find nonthermal effects result in $\tau_E \approx 2.5 \gamma_{ep}^{-0.75} \beta_{ee}^{-0.25}$. Here $\gamma_{ep}$ and $\beta_{ee}$ are measures of e-p and e-e interaction strength. $\gamma_{ep}$ is the two-temperature model prediction for the energy relaxation rate [29]. $\beta_{ee}$ is the electron-electron relaxation rate for an electron/hole 0.5 eV above/below the Fermi level. We find that the energy relaxation time $\tau_E$ remains sensitive to e-e scattering unless $\tau_E$ is at least two orders of magnitude larger than $\tau_H$. Alternatively, we find the dependence of $\tau_H$ on e-e versus e-p interactions is quite different than for $\tau_E$. We find that in most cases, due to differences in the nature of e-e vs. e-p interactions, the time-scale for high energy electrons to relax, $\tau_H$, will depend primarily on e-e interactions. For photoexcitation with $hv \geq 2$ eV, $\tau_H$ depends only on e-e quasi-particle lifetimes. This is true even if e-p quasiparticle lifetimes, $\tau_{ep}$, are hundreds of times shorter than e-e quasi-particle lifetimes, $\tau_{ee}$. In order for $\tau_H$ to be sensitive to e-p scattering rates, $hv$ needs to be in the near-infrared, e.g. ~1 eV, and $\gamma_{ep} / \beta_{ee}$ must be larger than 1. $\gamma_{ep} / \beta_{ee}$ is larger than 1 for metals with light elements, e.g. Al, Cu, and Li. Our findings for $\tau_H$ agree with prior studies on nonequilibrium electron dynamics that found the lifetime of photoexcited electrons depends only on e-e quasi-particle lifetimes [1,24,33].

**Results**

**Equation of Motion for Nonequilibrium Electron Dynamics**

To accurately capture the interplaying effects of electron-electron and electron-phonon scattering on the dynamics, we solve the equation of motion for the electron distribution function in a simple metal

$$\frac{df(\varepsilon,t)}{dt} = \Gamma_{ee}(f(\varepsilon,t)) + \Gamma_{ep}(f(\varepsilon,t)) \quad (1)$$

Here $\varepsilon$ is electron's energy relative the Fermi-level, $\Gamma_{ee}$ is the e-e collision integral [28], and $\Gamma_{ep}$ is the e-p collision integral [29]. Equation (1) accounts for both increases and decreases in $f(\varepsilon,t)$ due to scattering events. As a result, the dynamics predicted by Eq. (1) are different from the simple exponential functions arrived at by applying the relaxation-time-approximation. Since we are interested in the time-evolution of the nonequilibrium electrons, we linearize Eq. (1) by defining the nonequilibrium distribution as $\phi(\varepsilon,t) = f(\varepsilon,t) - f_0(\varepsilon, T_p)$. Here $f_0$ is the thermal Fermi-Dirac and $T_p$ is the temperature of the lattice. Our use of the phrase nonequilibrium electrons, or hot electrons, refers to the electrons and holes described by $\phi(\varepsilon,t)$.

The two-temperature model is a special limit of Eq. (1). The two-temperature model assumes $f(\varepsilon,t)$ is described by Fermi-Dirac statistics with an electron temperature $T_e$ distinct from $T_p$. For this special limit [29], Eq. (1) reduces to a simple heat-equation

$$C_e \frac{\partial T_e}{\partial t} = g_{ep}[T_p - T_e]. \quad (2)$$

Here, $C_e$ is the electron heat-capacities and $g_{ep} = (\pi \hbar k_B D_F) \lambda \langle \omega^2 \rangle$. $D_F$ is the density of states at the Fermi level, and $\lambda \langle \omega^2 \rangle$ is the 2$^{nd}$ frequency moment of the e-p spectral function [29]. $\lambda \langle \omega^2 \rangle$ is a measure of the strength of e-p interactions at the Fermi-level (see Methods and Supplementary Notes 1 and 2). The dynamics of $T_p$ are typically described with a 2$^{nd}$ heat-equation for the phonon sub-system (not shown here). The two-temperature model predicts an energy relaxation rate of $\gamma_{ep} = g_{ep}\left(C_p^{-1} + C_e^{-1}\right)$ [52]. At room temperature, where $C_p \gg C_e$, this simplifies to $\gamma_{ep} \approx g_{ep}/C_e$. The two-temperature model energy relaxation rate depends only on the strength of e-p interactions in the metal, $\gamma_{ep} \approx 3\hbar \lambda \langle \omega^2 \rangle / (\pi k_B T)$ [29].

**Dynamics Depend on Quasi-particle Interaction Strengths**

To quantify the parametric dependence of $\tau_E$ and $\tau_H$ on the strength of both e-e and e-p interactions we need descriptors of the e-e and e-p interaction strengths. Somewhat arbitrarily, we choose $\gamma_{ep}$ and $\beta_{ee}$ as descriptors of the e-e and e-p interaction strength. $\gamma_{ep}^{-1}$ is the $\tau_E$ predicted by the two-temperature model [29], while $\beta_{ee}^{-1}$ is the e-e relaxation time for 0.5 eV excitations, $\beta_{ee} = \tau_{ee}^{-1}(\varepsilon = 0.5 \text{ eV})$. There are a variety of other physical properties that would serve equally well as descriptors. We discuss descriptor choice in more detail in Methods. In Table 1, we report literature values for $\gamma_{ep}$ and $\beta_{ee}$ for various metals.

We summarize the dynamics predicted by Eq. (1) in Fig. 2. Figure 2a shows the total number of nonequilibrium electrons vs. time for different ratios of e-p to e-e interaction strength $\gamma_{ep}/\beta_{ee}$. Figure 2b shows how the energy distribution of nonequilibrium electrons evolves with time. For realistic values of e-e interaction and e-p interaction strengths, e.g. $\gamma_{ep} / \beta_{ee} \approx 0.25$, e-e scattering increases the number of nonequilibrium electrons by about a factor of 5 on a $\tau_E$ time-scale. Alternatively, for $\gamma_{ee} / \beta_{ee} \to 0$ the energy stored in the initial nonthermal distribution instantly redistributes into a thermal distribution and Eq. (2) governs the dynamics. A thermalized electron distribution has ~16x as many exictations as are initially photo-excited. Approximately 90% of excitations in a thermal distribution are within ~100 meV of the Fermi-level. The difference between these two cases of realistic vs. infintiely strong e-e interactions is sometimes discussed in terms of a maximum equivalent effect temprature $\Delta T_e^{me}$. $\Delta T_e^{me}$ is defined as the temperature increase of a thermalized electron gas for the same injected energy [32,41,46].

In Supplementary Figures 1-3, we show dynamics for Pt, Au, and Al. Specifically, we show the time-evolution of the occupation vs. energy, $\phi(\varepsilon)$, and energy-distribution vs. energy, $\varepsilon\phi(\varepsilon)$. The metals Pt, Au, and Al were chosen to illustrate dynamics for metals with small, typical, and large values of $\gamma_{ep} / \beta_{ee}$ in Table 1, respectively. In Supplementary Movies 1 and 2, we show the time-evolution of $\phi(\varepsilon)$ for Au as a function of time passing on a linear and logarithmic rate, respectively.

Our results for $\phi(\varepsilon,t)$ yield dynamics like those reported in many prior studies that solved Eq. (1) without using the relaxation-time-approximation [5,6,19,25,30,32,40]. Prior studies of nonequilibrium dynamics that solve the collision integrals in Eq. (1) have focused on specific material systems, e.g. Al, Au, Cu, and Ag [19,25,30,32,40]. New to our study is explicit consideration of how dynamics evolve across a wide range of e-e and e-p scattering strengths. Our predictions for the dynamics of $\phi(\varepsilon)$ differ significantly from prior studies that incorrectly approximate Eq. (1) with a relaxation-time approximation [1,2,36,45,51]. Models that use relaxation-time type approximations will predict $\tau_H$ values that are too short, because they assume e-p interactions effect $\tau_H$. Furthermore, while relaxation-time models, like the modified two temperature model [53], assume the time-scale for a nonequilibrium electrons to thermalize is approximately equal to $\tau_H$, we find the time-scale for thermalization is comparable to $\tau_E$.

**Dependence of $\tau_H$ and $\tau_E$ on Quasi-particle Interaction Strengths**

From $\phi(\varepsilon,t)$ predicted by Eq. (1), we determine relaxation times $\tau_H$ and $\tau_E$ as a function of e-p and e-e interaction strengths. Figure 3 shows how $\tau_H$ (time for high energy electrons to decay into lower energy electrons) and $\tau_E$ (time for energy of the nonequilibrium electrons to be transferred to the lattice) depend on $\gamma_{ep} / \beta_{ee}$. Figure 3 is the primary result of our study. We observe that $\tau_H$ and $\tau_E$ possess a universal dependence on the ratio of e-p to e-e scattering strengths. We find that in nearly all metals, $\gamma_{ep} / \beta_{ee}$ is such that $\tau_H$ depends only on e-e, while $\tau_E$ is determined by both e-e and e-p. To illustrate this universal dependence, we report $\tau_E$ normalized by $\gamma_{ep}^{-1}$, and $\tau_H$ normalized by $\beta_{ee}^{-1}$. The slope of $\tau_E \gamma_{ep}$ vs. $\gamma_{ep} / \beta_{ee}$ is determined by the sensitivity of $\tau_E$ to e-e interactions. A slope of zero indicates that energy-exchange between electrons and phonons is not affected by the strength of e-e interactions. Similarly, the slope of $\tau_H \beta_{ee}$ vs. $\gamma_{ep} / \beta_{ee}$ is determined by the sensitivity of $\tau_H$ to e-p interactions. A slope of zero indicates the time for high energy electrons to decay into lower energy electrons is determined only by e-e interactions.

**Discussion**

We now discuss the origins for the dependence of $\tau_E$ and $\tau_H$ on $\beta_{ee}$ and $\gamma_{ep}$. For most metals, high energy electrons decay with $\tau_H \approx C\beta_{ee}^{-1}$, where $C \approx 0.8 \text{ eV}^2/(hv)^2$ with our model assumptions. In general, $C$ will depend on $\phi(\varepsilon, t=0)$ and the energy dependence of the e-e scattering times. $\tau_H$ depends solely on the e-e interaction strength for two reasons. First, e-e scattering causes much larger changes in the average energy per excitation than e-p interactions. For an electron at energy $\varepsilon = hv$, the most probable amount of energy exchanged in an e-e interaction is $hv$ [26]. Alternatively, an e-p interaction will, on average, change the electron's energy by $\hbar\langle\omega\rangle$. Here, $\hbar\langle\omega\rangle$ is the average phonon energy of the metal and is typically 50-100x smaller than the photon energy $hv$. The second reason e-e interactions dominate $\tau_H$ is related to the number of in vs. out scattering events for high energy excitations. Nearly all e-e scattering events relax high energy excitations, but only a fraction of e-p scattering events do the same. There are three types of e-p interactions in the e-p collision integral: spontaneous phonon emission, stimulated phonon emission, and phonon absorption. Phonon absorption and stimulated emission rates are nearly equal. Phonon absorption increases an electron's energy, while stimulated phonon emission decreases it. As a result, the most important e-p interaction for $\phi(\varepsilon,t)$ is spontaneous phonon emission. The net effect of all e-p interactions on dynamics is a decrease in energy per electron at a rate of $\pi^2 k_B T \gamma_{ep}/3$. If all e-p interactions reduced electron energies, the energy per electron would decrease at a faster rate of $\hbar\langle\omega\rangle\tau_{ep}^{-1}$.

While e-p interactions won't influence $\tau_H$ in metals, they will affect the momentum distribution of nonequilibrium electrons. For some phenomena, e.g. energy transport or photo-catalysis, the momentum distribution of nonequilibrium electrons is also important.

The relationship we observe in Fig. 3 of $\tau_H \approx C\beta_{ee}^{-1}$ breaks down in the limit of very strong e-p interactions, e.g. $\gamma_{ep}/\beta_{ee} \gg 1$, and/or for photon energies less than 1 eV. (Supplementary Note 3 provides a phenomonlogical expression for $\tau_H$ that works across a wider range of $\gamma_{ep}/\beta_{ee}$ values.) For low energy excitation, e.g. $h\nu < 1\,\text{eV}$, a significant percentage of initially excited carriers are within a few hundred meV of the Fermi-level, where e-p interactions dominate dynamics. In the limit $h\nu > 1\,\text{eV}$ and $\gamma_{ep}/\beta_{ee} \gg 1$, the product of $\tau_H$ and $\beta_{ee}$ is not constant, meaning $\tau_H$ depends on both e-e and e-p interaction strength. However, for metals where literature data is available for both $\gamma_{ep}$ and $\beta_{ee}$, we could find no examples where $\gamma_{ep}/\beta_{ee} \gg 1$. Metallic compounds with exceptionally strong e-p interactions, such as Be, VN and MgB$_2$ with $\lambda\langle\omega^2\rangle \approx 2000\,\text{meV}^2$, do not have data available for e-e lifetimes. If these metals possessed weak e-e interaction strengths, e.g. $\beta_{ee}^{-1} > 50\,\text{fs}$, then $\tau_H$ would be sensitive to the e-p interaction strength.

In contrast to $\tau_H$, $\tau_E$ is sensitive to both e-e and e-p scattering so long as $\gamma_{ep}/\beta_{ee} > 0.05$. While it is obvious the time-scale for energy transfer from electrons to phonons should depend on e-p scattering strength, the importance of e-e scattering is less straightforward. Unlike e-p scattering, e-e interactions do not change the total energy in the electronic subsystem. Instead, e-e interactions alter how energy is distribtued across the electronic subsystem. Electron-electron scattering events turn a single excited electron into three excited electrons. Three electrons will transfer energy to the phonons roughly three times as fast as one electron because they will spontaneously emit phonons three times as often. As a result, both e-e and e-p interactions determine $\tau_E$ if electronic interactions don't rapidly thermalize the electronic subsystem. The importance of cascade dynamics on energy-transfer rates was reported by Tas and Maris in 1994 [5] as well as others later [24,30,32,40].

The energy relaxation times in Figure 3 are well approximated as $\tau_E \approx 2.5 \cdot \beta_{ee}^{-0.25} \gamma_{ep}^{-0.75}$ provided $0.05 < \gamma_{ep} / \beta_{ee} < 2$. Alternatively, $\tau_E \approx \gamma_{ep}^{-1} + 1.8 \gamma_{ep}^{-1} \left[ 1 - \tanh\left(-0.35 \ln\left[ 0.6 \gamma_{ep} / \beta_{ee} \right]\right)\right]$ is a good approximation for all $\gamma_{ep} / \beta_{ee} < 2$. A survey of literature values for e-e and e-p interaction strength suggest nearly all metals fall in the range of $0.05 < \gamma_{ep} / \beta_{ee} < 2$, see Table 1. For these metals, the two-temperature model estimate of $\tau_E$ is off by a factor ranging from 1.3 to 3, depending on the ratio $\gamma_{ep} / \beta_{ee}$.

Nonthermal effects are most important in metals with light elements and simple electronic structures where $\gamma_{ep} / \beta_{ee}$ is largest, e.g. Al. $\gamma_{ep}$ is highest in metals with light elements, because small ion mass leads to higher phonon frequencies and stronger electron-phonon coupling. $\beta_{ee}$ is smallest in metals where phase-space for e-e scattering is limited, e.g. the noble metals. $\beta_{ee}$ is largest in transition metals where the Fermi-level lies in the d-bands. Partial occupation of d-bands increases the phase-space for e-e scattering processes by allowing interband transitions. All other factors being equal, $\beta_{ee}$ will be higher in metals with higher charge densities. Screening effects are also important. $\beta_{ee}$ is smaller in metals where screening is large (higher permittivity in the static limit.) For example, $\beta_{ee}$ is smaller in Au than Ag due to d-band screening.

In the limit of strong e-e scattering, $\gamma_{ep}/\beta_{ee} < 0.05$, the energy relaxation time converges to the two-temperature model prediction, $\tau_E \approx \gamma_{ep}^{-1}$. In this limit, the relaxation of the nonequilibrium electron distribution occurs in a two-step process. The first step is e-e scattering drives electrons into a distribution that is nearly thermal, i.e. a distribution that maximizes entropy in the electronic subsystem. At this stage, the electrons remain out-of-equilibrium with the phonons, i.e. $T_e \neq T_p$. The second step is the nonequilibrium electrons transfer energy to the lattice on a $\gamma_{ep}^{-1}$ time-scale. The alkali metals Na, K, Rb, and Cs have sufficiently weak e-p interactions for the two-temperature model to be valid. Pd and Pt are also close to meeting the $\gamma_{ep}/\beta_{ee} < 0.05$ criteria due to strong e-e interactions.

While the two-temperature model will lack predictive power in most systems made up of only one metal, $\gamma_{ep}/\beta_{ee} < 0.05$ is easier to satisfy in bilayer systems composed of different types of metals. In a bilayer, if one metal has strong e-e interactions, while the other has weak e-p interactions, e.g. Pt with Au [15,39], then photoexcited electrons in these systems will relax via a two-step process similar to the one described above for two-temperature behavior [15,17]. First nonequilibrium electrons will thermalize in the layer with strong e-e scattering. Second, a now thermalized distribution of nonequilibrium electrons will exchange energy with phonons in the metal layer with weak e-p interactions. Several recent experimental studies have observed two-step dynamics in metal bilayer systems [15,17,39].

Now we compare our model predictions for $\tau_E$ of Au, Al, and Pt with experiment. While a variety of experimental studies are sensitive to the cooling rates of photoexcited electrons [48], interpretation of such experiments is not straightforward [19,38,54]. Time-resolved measurements of changes in optical properties, e.g. time-domain thermoreflectance or time-domain transient absorption, are common methods for studying nonequilibrium electron dynamics [1,27,34,37,48,55]. Optical properties depend on the excited electron distribution in a complex way and deducing $\tau_E$ from decay-rates of thermoreflectance or transient absorption signals is not trivial [19]. Two recent experimental studies on nonequilibrium electron dynamics in Au account for this complexity by modeling of how the nonequilibrium electron distribution correlates to changes in the dielectric function of Au. Both studies conclude nonequilibrium electrons transfer energy to phonons on a 2-3 ps time-scale, in fair agreement with our model's prediction of $\tau_E \sim 2$ ps. Our model predictions for $\tau_E \approx 0.15$ ps in Al and $\tau_E \approx 0.16$ ps in Pt are shorter than experimental values extracted from measurements of nonequilibrium heat transfer in metals. Tas and Maris report $\tau_E \approx 0.23$ ps in Al[5], while Jang *et al.* report $\tau_E \approx 0.2$ ps for Pt [56].

The reasonable agreement between our prediction of $\tau_E \approx 0.16$ ps for Pt and the Jang *et al.* experimental value of $\tau_E \approx 0.2$ ps is likely coincidental because there could be error in the value of $\gamma_{ep}$ we use for Pt. The values of $\gamma_{ep}$ in Table 1 for all metals were determined in a crude manner based on Debye temperatures and an analysis of experimental electrical resistivity data[57]. Such an approach is likely to have significant error for a metal like Pt, where the electronic density of states is a strong function of energy near the Fermi-level[56].

The discrepancy between the experimental value for Al of $\tau_E \approx 0.23$ ps and our prediction of $\tau_E \approx 0.15$ ps is surprising. Al is often viewed as a nearly free electron metal. Our model assumptions should be most reasonable for free electron like metals. The discrepancy may be due to our estimate of $\beta_{ee}$. For Al we set $\beta_{ee}^{-1}$ to 40 fs for Al based on time-resolved two-photon photoemission data[33]. However, deducing the average e-e scattering rate vs. electron energy from experimental photoemission data is non-trivial. It requires evaluating the effect of a variety of factors on the on the two photon photoemission data. These factors include hot electron transport, surface scattering, e-p interactions, and the wave-vector dependence of e-e scattering rates [33]. For many metals, predictions for $\beta_{ee}^{-1}$ from the *GW* approximation [26] agree with photoemission data, e.g. Au (220 versus 300 fs) and Cu (200 versus 160 fs). But this is not the case for Al, where *GW* predicts a electron-momentum averaged value for $\beta_{ee}^{-1}$ that is ~6x larger than the one deduced from two-photon photoemission measurements [26]. Nechaev *et al.* have suggested the disagreement is because the two-photon data is a measure of both e-e and e-p interactions in Al[58]. However, Nechaev *et al.* analysis does not solve an equation of motion for hot electrons like Eq. (1) to include the effect of e-p interactions on the distribution. Instead, Nechaev et al.'s analysis relies on Matthiesen's rule to add e-e and e-p quasiparticle scattering rates, which is not valid. Schone *et al.* have suggested that two-photon photoemission experiments are primarily a measure of the lifetime of electrons near the W-point of k-space[59]. Light absorption primarily populates states near the W-point of k-space due to momentum conservation. Near the W-point, the band-structure of Al deviates markedly from a free-electron system[59]. As a result, the e-e quasi-particle lifetime of electronic states near the W-point are much shorter than the average value across the Brillouin zone [59]. Since the time-scale for energy relaxation is much greater than the time-scale for e-e and e-p quasi-particle scattering, $\tau_E$ will depend on e-e scattering rate of states across the entire

Brillouin zone, and not just e-e scattering rates of states near the W-point. If, instead of using two-photon data, we use the electron-momentum averaged *GW* prediction from Ladstädter *et al.*[26] to set $\beta_{ee}^{-1} \approx 260$ fs for Al, our model predicts $\tau_E \approx 0.22$ ps. This latter value is in good agreement with the experimental results of Tas and Maris[5].

While the present study considers the regime of low laser fluence, we expect that at larger fluence the type of dynamics, and relaxation times, will be different. At higher fluence, the dynamics will be closer to the two-step process described by the two-temperature model. This change in dynamics occurs because a higher laser fluence requires fewer e-e scattering events to relax photoexcited electrons to a Fermi-Dirac thermal distribution. To understand why, consider an absorbed fluence of 10 mJ m$^{-2}$ in a 10 nm thick Au film. This energy density spread across a thermal distribution of electrons corresponds to 60 meV per excited electron, much less than eV scale energies of photoexcited electrons. Alternatively, an absorbed fluence of 10 J m$^{-2}$ spread across a thermal distribution of electrons corresponds to ~0.5 eV per excited electron, which is comparable to the energy of photoexcited electrons. Therefore, a distribution excited by a high fluence laser pulse requires fewer e-e scattering events to evolve into a Fermi-Dirac distribution.

Our calculations in Figs. 1-3 were carried out at 300 K, but the results are similar at other temperatures. The rate of energy relaxation will increase at lower temperatures because of decreases in electronic heat capacity, i.e. changes in $f_0(\varepsilon, T)$. Changes to e-p scattering rates due to changes in ambient temperature are relatively unimportant. This is because the rate of energy transfer from nonequilibrium electrons to phonons depends primarily on spontaneous phonon emission, which is temperature independent. The effect of temperature is included in our approximate expression for $\tau_E$ via the $\gamma_{ep}$ term.

In conclusion, we have numerically solved the Boltzmann rate equation to quantify how cascade dynamics of photoexcited electrons depend on e-e and e-p interactions. For most simple metals, the rate of energy transfer is sensitive to both e-e scattering and e-p scattering due to cascade dynamics. We find nonthermal effects are most important in metals with light elements and simple electronic structures, e.g. Al and Li. The energy relaxation time of the nonequilibrium electron distribution is well approximated as $\tau_E \approx 2.5 \cdot \beta_{ee}^{-0.25} \gamma_{ep}^{-0.75}$, where $\gamma_{ep}$ is the electron-phonon energy relaxation rate predicted for a thermal electron distribution, and $\beta_{ee}$ is e-e scattering rate of an electron or hole 0.5 eV away from the Fermi level. In the limit that $\gamma_{ep}/\beta_{ee} < 0.05$, the two-temperature model is accurate because e-e scattering is effective at establishing a near thermal distribution of electrons before significant energy is transferred to the lattice. We can identify only a few metals that satisfy the criterion $\gamma_{ep}/\beta_{ee} < 0.05$: Na, K, Rb and Cs. These findings are important for understanding ultrafast electron dynamics in a diverse range of fields, e.g. ultrafast magnetism, photocatalysis, plasmonics, and others.

**Methods**

**Collision Integrals.** To solve Eq. (1) we need analytic expressions for the collision integrals. Using a Taylor series expansion, we approximate the electron-phonon collision integral as

$$\Gamma_{\text{ep}}(\phi(\varepsilon,t)) = \pi\hbar\lambda\langle\omega^2\rangle\left[-2\frac{df_0(\varepsilon)}{d\varepsilon}\phi(\varepsilon) + [1-2f_0(\varepsilon)]\frac{\partial\phi(\varepsilon)}{\partial\varepsilon} + k_B T\frac{\partial^2\phi(\varepsilon)}{\partial\varepsilon^2}\right]. \tag{3}$$

Here, $\lambda\langle\omega^2\rangle$ is the second frequency moment of the Eliashberg function $\alpha^2 F(\omega)\omega^{-1}$,

$$\lambda\langle\omega^2\rangle = 2\int d\omega\, \alpha^2 F(\omega)\omega. \tag{4}$$

We provide a full derivation of Eq. (3) in Supplementary Note 1. We use the analytic solution for the electron-electron collision integral derived by Kabanov *et al.*[28] for Fermi liquids

$$\frac{d\phi}{dt} = -\frac{\phi(\varepsilon)}{\tau_{\text{ee}}(\varepsilon)} + \frac{K}{\cosh\left(\frac{\varepsilon}{2k_B T}\right)}\int_{-\infty}^{\infty} d\varepsilon'\phi(\varepsilon')\cosh\left(\frac{\varepsilon'}{2k_B T}\right)$$

$$\times\left[\frac{(\varepsilon-\varepsilon')}{\sinh\left(\frac{\varepsilon-\varepsilon'}{2k_B T}\right)} - \frac{(\varepsilon+\varepsilon')}{2\sinh\left(\frac{\varepsilon+\varepsilon'}{2k_B T}\right)}\right] \tag{5}$$

where

$$\tau_{\text{ee}}(\varepsilon) = \frac{2}{K}\left(\frac{1}{(\pi k_B T)^2 + \varepsilon^2}\right). \tag{6}$$

Eq. (5) redistributes energy in the electronic subsystem while conserving the total energy in the subsystem.

Evaluation of $\Gamma_{ep}$ in Eq. (3) requires the material's electron-phonon spectral function $\alpha^2 F(\varepsilon, \omega)$ [24,25,28,30,32] to evaluate the value of $\lambda \langle \omega^2 \rangle$ as a function of electron energy $\varepsilon$. Similarly, evaluation of $\Gamma_{ee}$ in Eq. (5) requires knowledge of the Kernel function $K(\varepsilon, \varepsilon', \varepsilon'', \varepsilon''')$ [24,25,28,30,32]. (This function is the Kernel of the e-e collision integral.) The function $\alpha^2 F(\varepsilon, \omega)$ is the average square of the electron-phonon matrix element on a constant electron energy surface of $\varepsilon$ with phonons of frequency $\omega$. The function $\alpha^2 F(\varepsilon, \omega)$ determines average e-p quasi-particle lifetime of electronic states on the constant energy surface $\varepsilon$ with a phonon of frequency $\omega$. At the Fermi-level, $\alpha^2 F(0, \omega)$ governs many electronic phenomena in metals, e.g. electrical resistivity and superconductivity [60]. The Kernel function $K(\varepsilon, \varepsilon', \varepsilon'', \varepsilon''')$ is the average square of the electron-electron matrix element between electrons on a constant energy surface $\varepsilon$ with electronic states on constant energy surfaces defined by $\varepsilon'$, $\varepsilon''$, and $\varepsilon'''$ [28]. The Kernel function determines the average e-e quasi-particle lifetime of electronic states on the constant energy surface $\varepsilon$. Like the e-p spectral function, the Kernel function is important for a variety of electronic phenomena in metals. At the Fermi-level, the constant $K(\varepsilon = 0)$ is related to the Coulomb pseudopotential, which is an important for the theory for low-temperature resistivity of transition metals [61] and the theory of superconductivity [28].

To define simple descriptors for the e-e and e-p interaction strengths, we neglect the dependence of $\lambda \langle \omega^2 \rangle$ and $K$ on electron energy $\varepsilon$ and fix e-e and e-p interaction strengths to their values at the Fermi-level. This assumption is quite good for simple metals like Al, Cu, Ag, Au, as well as the alkali metals. In these metals, the electronic density of states is relatively constant within $h\nu$ of the Fermi-level, which is the energy-scale we are concerned with here. A variety of theoretical and experimental studies provide evidence that neglecting the $\varepsilon$ dependence is a reasonable approximation for simple metals. First-principles calculations for Al, Cu, and Au confirm that both $\lambda\langle\omega^2\rangle$ [18] and $K$ [26] depend only weakly on $\varepsilon$. An energy independent $K$ leads to the well-known $\varepsilon^{-2}$ dependence for electron-hole excitations in a Fermi-liquid, and time-resolved two-photon photoemission measurements of Al, Au, Ag, and Cu observe such an $\varepsilon^{-2}$ dependence [33]. Alternatively, in transition metals, the electronic density of states can vary significantly within a few eV of the Fermi level. As a result, our assumption that $\lambda\langle\omega^2\rangle$ and $K$ are independent of $\varepsilon$ will cause some error in calculated values of $\tau_E$ and $\tau_H$ for transition metals. We quantify this error in Supplementary Note 2.

Instead of using $\lambda\langle\omega^2\rangle$ and $K$ as descriptors for the e-e and e-p interaction strengths, we prefer alternative but related parameters that correspond to important time-scales in our problem. As a descriptor of the e-p interaction, we choose the energy-relaxation time for a thermal distribution of nonequilibrium electrons, $\gamma_{ep}$. $\gamma_{ep}$ and $\lambda\langle\omega^2\rangle$ are proportional to one another: $\gamma_{ep} = 3\hbar\lambda\langle\omega^2\rangle/(\pi k_B T)$. The values we used for $\gamma_{ep}$ of various metals are reported in Table 1. Table 1 values are based on an analysis of electrical resistivity data by Allen[60] and uses the approximation $\lambda\langle\omega^2\rangle \approx \lambda \cdot \Theta_D^2/2$, where $\Theta_D$ is the Debye temperature. To describe the e-e interaction strength, we choose the scattering rate of a 0.5 eV electronic excitation $\beta_{ee} = K(0.5eV)^2/2$. Values for $\beta_{ee}$ in Table 1 for non-alkali metals are based on photoemission data for electron lifetimes [33]. $\beta_{ee}$ for the alkali metals is based on predictions of Fermi liquid theory for a homogenous electron gas [33]. We choose the scattering time for 0.5 eV electrons as our measure for e-e interaction strength because this is the lowest energy where experimental two-photon emission data is commonly available. Alternative descriptor choices for e-e interactions, e.g. the lifetime of 1 eV electrons, would yield quite similar results (see Supplementary Note 2). Fixing the e-e interaction strength with the electron lifetime at 0.5 eV allows Eq. (6) to make reasonably accurate predictions for $\tau_{ee}(\varepsilon < 1eV)$ in transistion metals, despite our model neglecting the $\varepsilon$ dependence of $K$, see Supplementary Fig. 4. We want $\tau_{ee}(\varepsilon)$ be accurate for low energy excitations because, as shown in Fig. 2b, nearly all nonequilibrium electrons are at low energies on time-scales comparable to $\tau_E$.

Solving Eq. (1) requires initial conditions. We assume the probability a photon with energy $hv$ will move an electron from a state with energy $\varepsilon$ to a state with energy $\varepsilon + hv$ is proportional to $f_0(\varepsilon)(1 - f_0(\varepsilon + hv))$. This assumption results in a flat initial distribution of electrons and holes with concentration $\phi_0 \ll 1$ that extends to an energy $hv$ above and below the Fermi level. We consider $hv$ between 1 and 3 eV, i.e. visible light. We focus on visible light because most experimental studies on ultrafast electron dynamics use visible light for photoexcitation. Our conclusions do not rely on the assumption that a flat distribution is excited. We obtain similar results if we assume a completely different energy dependence for the initial distribution. For example, we obtain nearly identical results for how $\tau_E$ depends on e-e and e-p scattering strengths if we instead assume photons with energy $h\nu$ only excite electrons and holes at energy $h\nu/2$ above and below the Fermi level.

In our calculations, we assume instantaneous photoexcitation so the relaxation times of $\phi(\varepsilon,t)$ depend only on e-e and e-p interactions. The time-scales $\tau_E$ and $\tau_H$ describe the intrinsic response times of the metal, and do not depend on pulse duration of the photoexcitation. The effect of a finite pulse duration could be included in several ways. A time-dependent source term could be added to Eq. (1). Or, our solution for $\phi(\varepsilon,t)$ in response to initial conditions could be used to construct a Green's function solution to the problem.

**Model Assumptions.** For completeness, we now summarize all the assumptions in our model. Equation 1 assumes the distribution function depends only on energy and time, thereby neglecting variation in angles of the wavevector. When solving Eq. 1, we neglect any rise in internal energy of the lattice, i.e. we assume $T_p$ is constant. This assumption is reasonable because the phonon heat-capacity is large compared to the electron heat capacity. Furthermore, allowing $T_p$ to evolve with time wouldn't affect predictions for $\tau_H$ and $\tau_E$ because $T_p$ doesn't affect the two most important types of scattering processes: e-e scattering rates and spontaneous phonon emission rates. For some applications, e.g. photocatalysis, the increase in $T_p$ is important to track so that thermal and nonequilibrium electron phenomena can be differentiated [22,23]. The effects of a dynamic phonon temperature can be added to our model by solving the equation $C_p \frac{\partial T_p}{\partial t} = -\frac{\partial E_{tot}}{\partial t}$ simultaneously with Eq. (1), where $E_{tot}$ is the energy stored by the nonequilibrium electron distribution. Another assumption we make when solving Eq. (1) is low fluence photoexcitation.

We linearize Eq. (1) by assuming $\phi(\varepsilon,t) = f(\varepsilon,t) - f_0(\varepsilon) \ll 1$, and keeping only terms linear in $\phi(\varepsilon,t)$. As noted above, we neglect the dependence of the e-p spectral function on electron energy, and the dependence of the e-e Kernel function on electron energy. Finally, by setting the initial distribution to $\phi(\varepsilon,t=0) = \phi_0$ at all energies within $h\nu$ of the Fermi-level, we are assuming an energy independent joint density of states. These latter three assumptions are all related to the energy dependence of the electronic density of states. We discuss why these latter three assumptions are reasonable in Supplementary Note 2.

## Acknowledgements

The work by R. Wilson was supported by the U.S. Army Research Laboratory and the U.S. Army Research Office under contract/grant number W911NF-18-1-0364. The work by S. Coh was supported by NSF DMR-1848074.

## Author Contributions

R.B.W. developed the model. R.B.W. and S.C. designed the study, performed calculations, and wrote the manuscript.

## Competing Interests

We declare that none of the authors have competing financial or non-financial interests as defined by Nature Research.

## Data Availability

The datasets generated during and/or analysed during the current study are available from the corresponding author on reasonable request
26

**Figures**

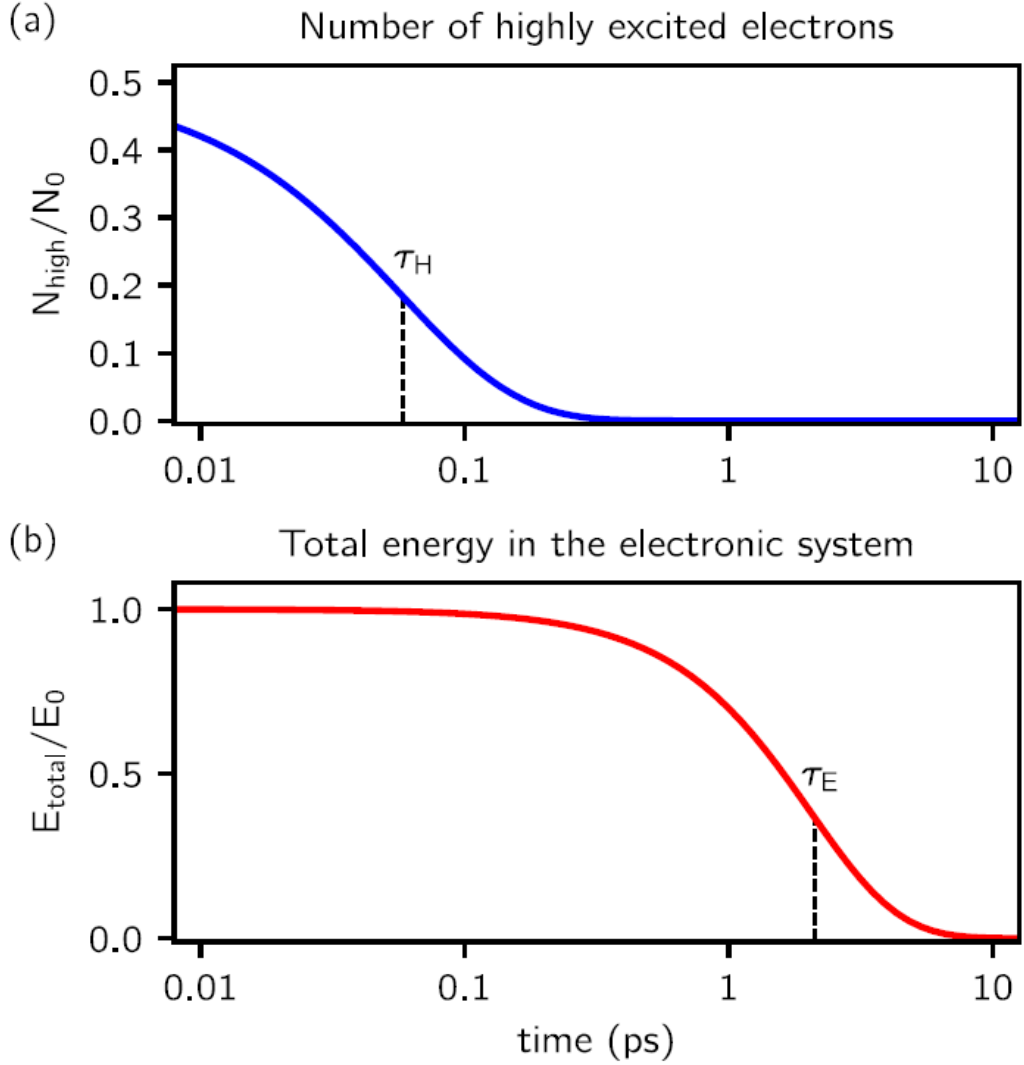

**Figure 1. Definitions of distribution relaxation time-scales $\tau_H$ and $\tau_E$.** We define time-scales $\tau_H$ and $\tau_E$ to characterize two distinct effects of quasi-particle interactions on nonequilibrium electron dynamics. $\tau_H$ measures how quickly electron-electron and electron-phonon interactions redistribute energy from high to low energy electronic states. $\tau_E$ measures how quickly electron-electron and electron-phonon interactions cause energy transfer from the electronic subsystem to the lattice. (a) After excitation with energy $hv$, the occupation states where $|\varepsilon| \geq hv/2$ decays with time $\tau_H$. Here, we show $\tau_H$ for Au. (b) The energy absorbed by the electrons remains in the electronic subsystem for time $\tau_E$. In Au, $\tau_E$ is 35 times greater than $\tau_H$.

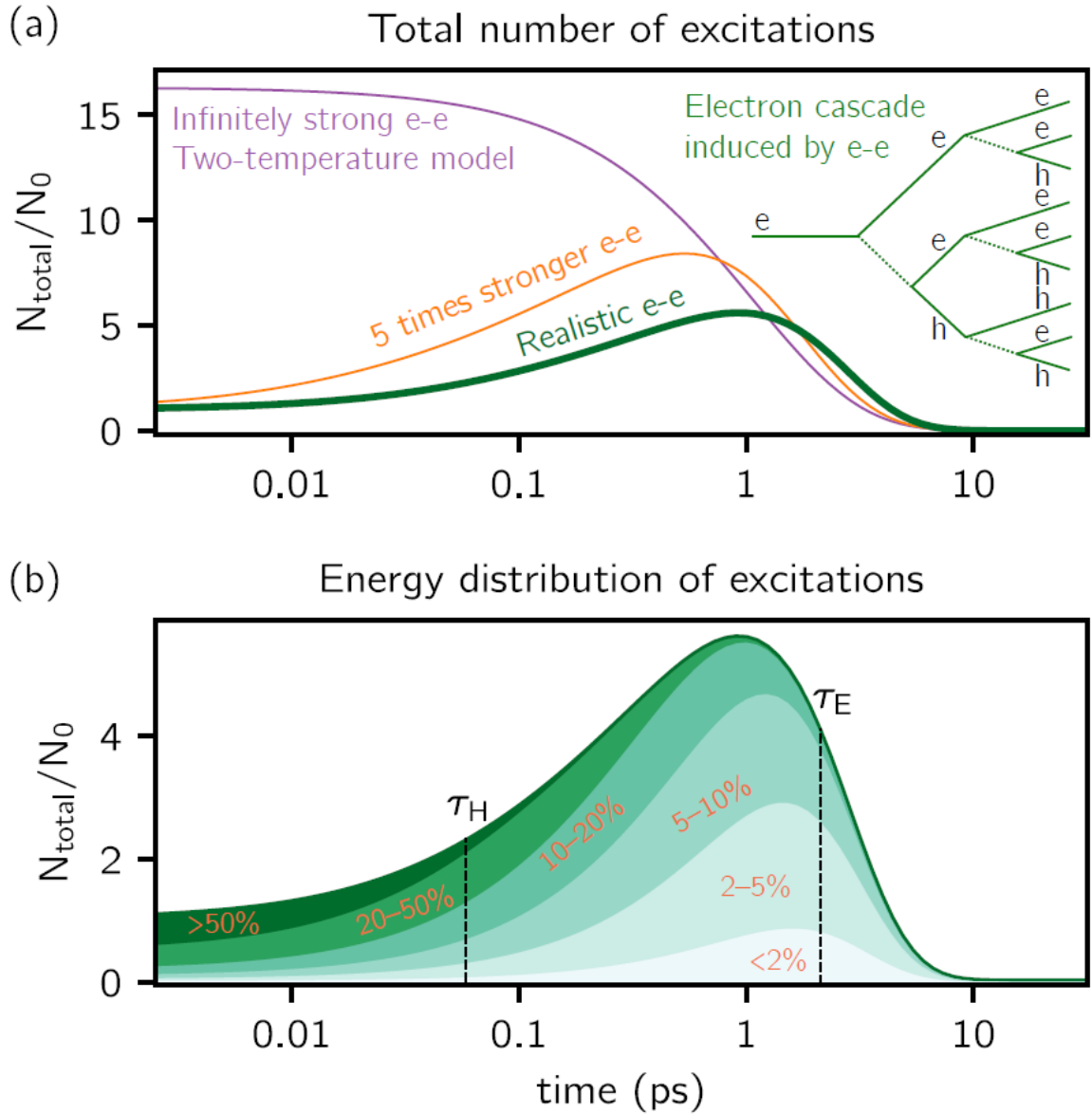

**Figure 2. Dynamics of Nonequilibrium Electrons after Photoexcitation with $h\nu = 2$ eV.**
(a) The total number of nonequilibrium electrons versus time for three different values of electron-electron (e-e) scattering strengths, $\gamma_{ep}/\beta_{ee} \approx 0.25$ (realistic e-e), 0.05 (strong e-e), and 0 (infinite e-e). For the case of infinitely strong electron-electron scattering, the initial distribution evolves instantaneously into a thermal distribution, which increases the number of hot electrons by a factor of ~16. The inset illustrates the cascade dynamics of nonequilibrium electrons, e, and nonequilibrium holes, h. (b) The energy distribution of excitations for the case of $\gamma_{ep}/\beta_{ee} \approx 0.25$. Each band represents the number of excitations in a specific energy range, e.g. the number of excitations with energy greater than 50% of $h\nu$ for the top most dark green band. $\tau_H$ is the time-scale that high energy electronic states remain occupied. $\tau_E$ is the time-scale for energy transfer between the electronic subsystem and lattice.

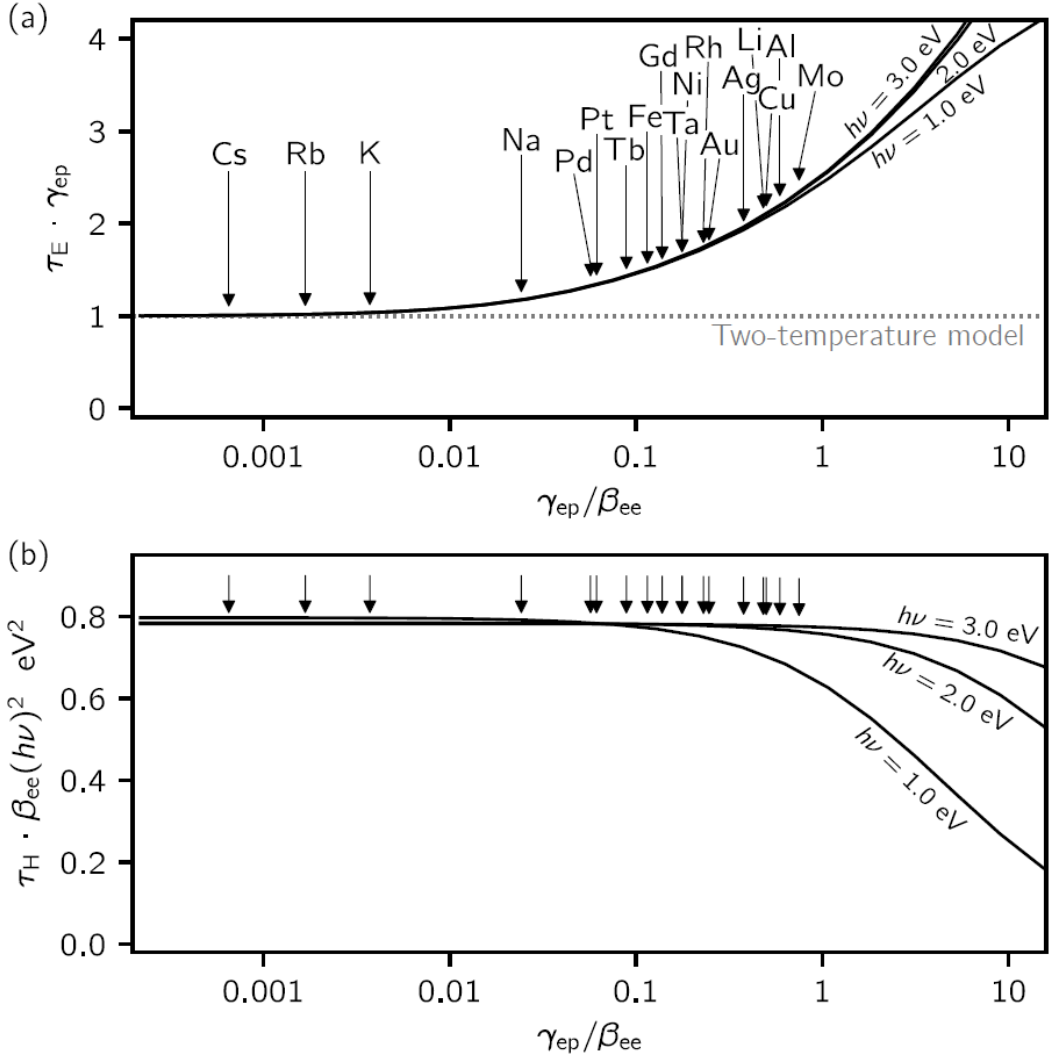

**Figure 3. Dependence of $\tau_H$ and $\tau_E$ on electron-electron and electron-phonon interaction strengths, $\beta_{ee}$ and $\gamma_{ep}$.** To illustrate the universal dependence of $\tau_E$ and $\tau_H$ on the ratio of interaction strengths $\gamma_{ep}/\beta_{ee}$, we report $\tau_E$ and $\tau_H$ normalized by the time-scales $\gamma_{ep}^{-1}$ and $\beta_{ee}^{-1}$, respectively. $\tau_H$ is the time-scale that high energy electronic states remain occupied. $\tau_E$ is the time-scale for energy transfer between the electronic subsystem and lattice. $h\nu$ is the energy of absorbed photons. Values of $\gamma_{ep}/\beta_{ee}$ for various metals are indicated with vertical arrows. (a) For realistic values of e-e vs. e-p interaction strengths, $\tau_E$ depends on both $\gamma_{ep}$ and $\beta_{ee}$. In the limit of $\gamma_{ep}/\beta_{ee} < 0.05$, $\tau_E$ converges to the two-temperature value and is independent of $\beta_{ee}$. (b) For photoexcitation with visible light (2 and 3 eV) and realistic values of e-e versus e-p interaction strengths, $\tau_H$ depends only on e-e interaction strengths.

**Table 1. Literature values for the electron-electron and electron-phonon interaction strengths, $\beta_{ee}$ and $\gamma_{ep}$, of various metals.** The table also shows calculated values for $\tau_H$, the time-scale that high energy electronic states remain occupied, and $\tau_E$, the time-scale for energy transfer between the electronic subsystem and lattice. The values for the second frequency moment of the Eliashberg function $\lambda\langle\omega^2\rangle$ and Debye temperature $\Theta_D$ are from Allen[60], Kittel[62], and Papaconstantopoulos et al.[63]. To highlight the large discrepancy between electron-phonon quasi-particle scattering time $\tau_{ep}$ and time-scales $\tau_H$ and $\tau_E$, we show $\tau_{ep} \approx \hbar/(2\pi\lambda k_B T)$ for each metal. However, we emphasize that $\tau_{ep}$ is not an input into our model. The values for $\beta_{ee}^{-1}$ for the alkali metals are predictions from Fermi-liquid theory for a homogenous electron gas[33]. The values for $\beta_{ee}^{-1}$ of other metals are from two photon photoemission data[33], except for Pt. We assume $\beta_{ee}^{-1}$ for Pt is equal to $\beta_{ee}^{-1}$ for Pd.

| Metal | $\lambda\langle\omega^2\rangle$ (meV$^2$/$\hbar^2$) | $\Theta_D$ (K) | $\tau_{ep}$ (fs) | $\gamma_{ep}^{-1}$ (fs) | $\beta_{ee}^{-1}$ (fs) | $\tau_E$ (fs) | $\tau_H$ (fs) |
|---|---|---|---|---|---|---|---|
| Li | 160 | 340 | 12 | 110 | 55 | 230 | 11 |
| Na | 13 | 158 | 29 | 1400 | 34 | 1600 | 7 |
| K | 3.4 | 91 | 37 | 5200 | 20 | 5500 | 4 |
| Rb | 1.8 | 56 | 27 | 9900 | 17 | 1.0x10$^4$ | 3 |
| Cs | 0.85 | 38 | 25 | 2.1x 10$^4$ | 14 | 2.1x10$^4$ | 3 |
| Ta | 190 | 240 | 4.6 | 93 | 17 | 150 | |
| Mo | 240 | 450 | 13 | 74 | 57 | 170 | |
| Fe | 280 | 470 | 12 | 63 | 7.5 | 92 | |
| Rh | 350 | 480 | 10 | 51 | 12 | 89 | |
| Ni | 230 | 450 | 13 | 77 | 14 | 120 | |
| Pd | 130 | 270 | 8.6 | 140 | 8 | 170 | |
| Pt | 140 | 240 | 6.1 | 100 | 8 | 160 | |
| Cu | 57 | 340 | 31 | 310 | 160 | 650 | 30 |
| Ag | 23 | 230 | 34 | 790 | 300 | 1500 | 60 |
| Au | 15 | 170 | 27 | 1200 | 300 | 2100 | 60 |
| Al | 270 | 430 | 10 | 67 | 40 | 150 | 8 |
| Gd | 90 | 200 | 7 | 200 | 28 | 290 | |
| Tb | 90 | 200 | 7 | 200 | 18 | 270 | |